\documentclass[11pt,a4paper,twocolumn]{article}
 
\usepackage[margin=1in]{geometry}
\usepackage{amsmath}
\usepackage{amssymb}
\usepackage{graphicx}
\usepackage{capt-of}
\usepackage{bm}
\usepackage{mathtools}
\usepackage{authblk}
\usepackage{cite}
\usepackage{hyperref}
\usepackage{multibib}
\newcites{supp}{References}

\title{\Large \vspace{-2\baselineskip} Generalized Skyrmions}
\date{}
\author[1,*]{An Aloysius Wang}
\author[1]{Zimo Zhao}
\author[1]{Yifei Ma}
\author[1]{Yuxi Cai}
\author[1]{Stephen M Morris}
\author[2]{Honghui He}
\author[3]{Lin Luo}
\author[4]{Zhenwei Xie}
\author[4]{Peng Shi}
\author[5]{Yijie Shen}
\author[6]{Anatoly V Zayats}
\author[4]{Xiaocong Yuan}
\author[1,*]{Chao He\vspace{-0.5\baselineskip}}
\affil[1]{\footnotesize Department of Engineering Science, University of Oxford, Parks Road, Oxford, OX1 3PJ, UK}
\affil[2]{\footnotesize Guangdong Research Center of Polarization Imaging and Measurement Engineering Technology, Tsinghua Shenzhen International Graduate School, Tsinghua University, Shenzhen 518055, China}
\affil[3]{\footnotesize College of Engineering, Peking University, Beijing 100871, China}
\affil[4]{\footnotesize Nanophotonics Research Center, Shenzhen Key Laboratory of Micro-Scale Optical Information Technology \& Institute of Microscale Optoelectronics, Shenzhen University, Shenzhen 518060, China}
\affil[5]{\footnotesize Division of Physics and Applied Physics, School of Physical and Mathematical Sciences,
Nanyang Technological University, Singapore, Singapore}
\affil[6]{\footnotesize Department of Physics and London Centre for Nanotechnology, King’s College London, London, UK}
\affil[*]{Corresponding authors: an.wang@stcatz.ox.ac.uk, aw6609@princeton.edu; chao.he@eng.ox.ac.uk\vspace{-2\baselineskip}}

\renewcommand{\figurename}{Fig.}

\begin{document}
\maketitle
{\bf Skyrmions are important topologically non-trivial fields characteristic of models spanning scales from the microscopic \cite{al_khawaja_skyrmions_2001, Fert2017} to the cosmological \cite{durrer_cosmic_2002}. However, the Skyrmion number can only be defined for fields with specific boundary conditions, limiting its use in broader contexts. Here, we address this issue through a generalized notion of the Skyrmion derived from the De Rham cohomology of compactly supported forms \cite{weintraub_differential_2014}. This allows for the definition of an entirely new $\coprod_{i=1}^\infty \mathbb{Z}^i$-valued topological number that assigns a tuple of integers $(a_1, \ldots, a_k)\in \mathbb{Z}^k$ to a field instead of a single number, with no restrictions to its boundary. The notion of the generalized Skyrmion presented in this paper is completely abstract and can be applied to vector fields in any discipline, not unlike index theory within dynamical systems \cite{strogatz2015}. To demonstrate the power of our new formalism, we focus on the propagation of optical polarization fields and show that our newly defined generalized Skyrmion number significantly increases the dimension of data that can be stored within the field while also demonstrating strong robustness. Our work represents a fundamental paradigm shift away from the study of fields with natural topological character to engineered fields that can be artificially embedded with topological structures.}

The intersection of topology and fields has given rise to some of the most fruitful developments in both modern mathematics and modern physics. These include applications of gauge theory to topology inspired by Donaldson's Theorem \cite{donaldson1983application} and the study of topological solitons arising from field equations, which includes the non-linear sigma models considered by Tony Skyrme \cite{skyrme1961non} from which the Skyrmion derives its name. 

A central theme underlying these developments, particularly in physics, is the idea of assigning fields a topological number. Through this assignment, an ostensibly continuous field structure can be given discrete and particle-like characteristics and, therefore, a notion of topological stability. 

It is not surprising, then, that the discovery of a new topological number is often the central catalyst for advancements across multiple scientific disciplines, as evidenced by the appearance of Skyrmions in diverse fields ranging from the study of Bose-Einstein condensates \cite{al_khawaja_skyrmions_2001} and magnetic spin textures \cite{Fert2017,Nagaosa2013} in condensed matter physics to optics and photonics \cite{Tsesses2018, shen_optical_2023, He2022, Shen2021, Shen2021_Super, Shen2023, Bai2020, Lin2021, he2023universal, du_deep-subwavelength_2019, lei_photonic_2021, shi_spin_2021, teng_physical_2023, wang2024unlock, Cisowski2023, mcwilliam_topological_2023, shi_strong_2020, Gao2020, ye2024theory, Liu2022}, cosmology \cite{durrer_cosmic_2002}, fluid dynamics \cite{PhysRevLett.132.054003}, soft matter physics \cite{foster_two-dimensional_2019}, acoustics \cite{acoustic, Hurtado_acoustic}, and more \cite{Cao_Skyrmion, yang_scalar_2023, ornelas_non-local_2024}. 

The Skyrmion, as we understand it today, derives its topological number from the degree of a smooth map originating from the De Rham cohomology \cite{naber_topology_2011}. This degree can be defined for smooth maps between compact, connected, orientable, smooth, $n$-dimensional manifolds, and is intuitively the number of times one manifold is ``wrapped'' around the other. However, as fields defined in real space do not naturally originate from such a manifold, work must be done to non-trivialize the domain of a physical field in order to define a topological number through the degree. This amounts to requiring that the field have appropriate symmetries on its boundary that allow one to ``glue'' the domain into a compact, connected, orientable, smooth, $n$-dimensional manifold \cite{wang2024topological}. The topological number of a field $\mathcal{S}\colon \Omega \longrightarrow Y$ compactifiable in this way can then be retrieved via an uncountable number of non-trivial integral equations,
\begin{equation}
    \deg \mathcal{S} = \int_{\Omega} \mathcal{S}^\ast\omega, \label{eq:Skyrmion Number}
\end{equation}
one for each normalized $\omega \in \Lambda^n(Y)$, of which the usual Skyrmion number integral equation is a special case.

Many important physical fields, such as magnetic spin textures \cite{Fert2017, Nagaosa2013}, naturally have the appropriate boundary conditions for compactification. More recently, developments in structured light have also enabled the synthesis of optical polarization fields with the necessary symmetries, such as through periodic evanescent fields \cite{Tsesses2018} and the superposition of specific Laguerre-Gaussian (LG) modes \cite{Gao2020, ye2024theory}. These discoveries have led to a growing interest in the study of optical Skyrmions and their applications to metrology, microscopy, sensing, communications, computing, and more \cite{shen_optical_2023}. 

Compared to more conventional Skyrmions, such as those observed in magnetic spin textures \cite{Fert2017, Nagaosa2013}, the optical Skyrmion is unique in many ways, the most significant of which is the capacity of optical fields to propagate. This property gives rise to the possibility of transporting topological information through electromagnetic fields, which has clear implications for high-density data applications. 

However, whether and how far the topological structure of an optical Skyrmion propagates remains a central yet challenging consideration in translating theory into real-world applications \cite{shi2024embarking}.

The lynchpin in establishing topological protection is the existence of an integral equation that is guaranteed to be integer-valued on every transverse plane of an optical polarization field as it propagates (Methods 1). From our earlier discussion, this is certainly true of fields that are compactifiable on every transverse plane, in which case the usual Skyrmion number integral suffices. However, compactifiability is a restrictive condition that excludes all but a narrow class of fields from consideration. To make matters worse, it is also not a property that is naturally preserved in propagation. 

The critical question then becomes: \\

\noindent {\it Without compactifiability, can we still find an integral equation that is guaranteed to be integer-valued on every transverse plane of an optical polarization field in propagation?}\\

The theory of generalized Skyrmions presented in this paper comes from our attempt to answer this question. In Methods 2, we show that for non-compactifiable fields, variations of the usual Skyrmion number integral derived through equation (\ref{eq:Skyrmion Number}) can be guaranteed to be integer-valued. However, unlike compactifiable fields where equation (\ref{eq:Skyrmion Number}) always evaluates to the same number, these modified integral equations can evaluate to different integers and define a tuple of numbers $(a_1, \ldots, a_k)\in \mathbb{Z}^k$ rather than a single one. This then allows for the definition of an entirely new topological number taking values in $\coprod_{i=1}^\infty \mathbb{Z}^i$ that is significantly more well-behaved in propagation compared to the usual Skyrmion number (see Fig.\ \ref{fig: Concept1}, which demonstrates this fact for three important situations, including propagation in free space, through a spatially varying birefringent medium and in a cylindrical conducting waveguide). 

\vspace{-\baselineskip}
\begin{center}
\begin{figure}[!ht]
\includegraphics[width=0.49\textwidth]{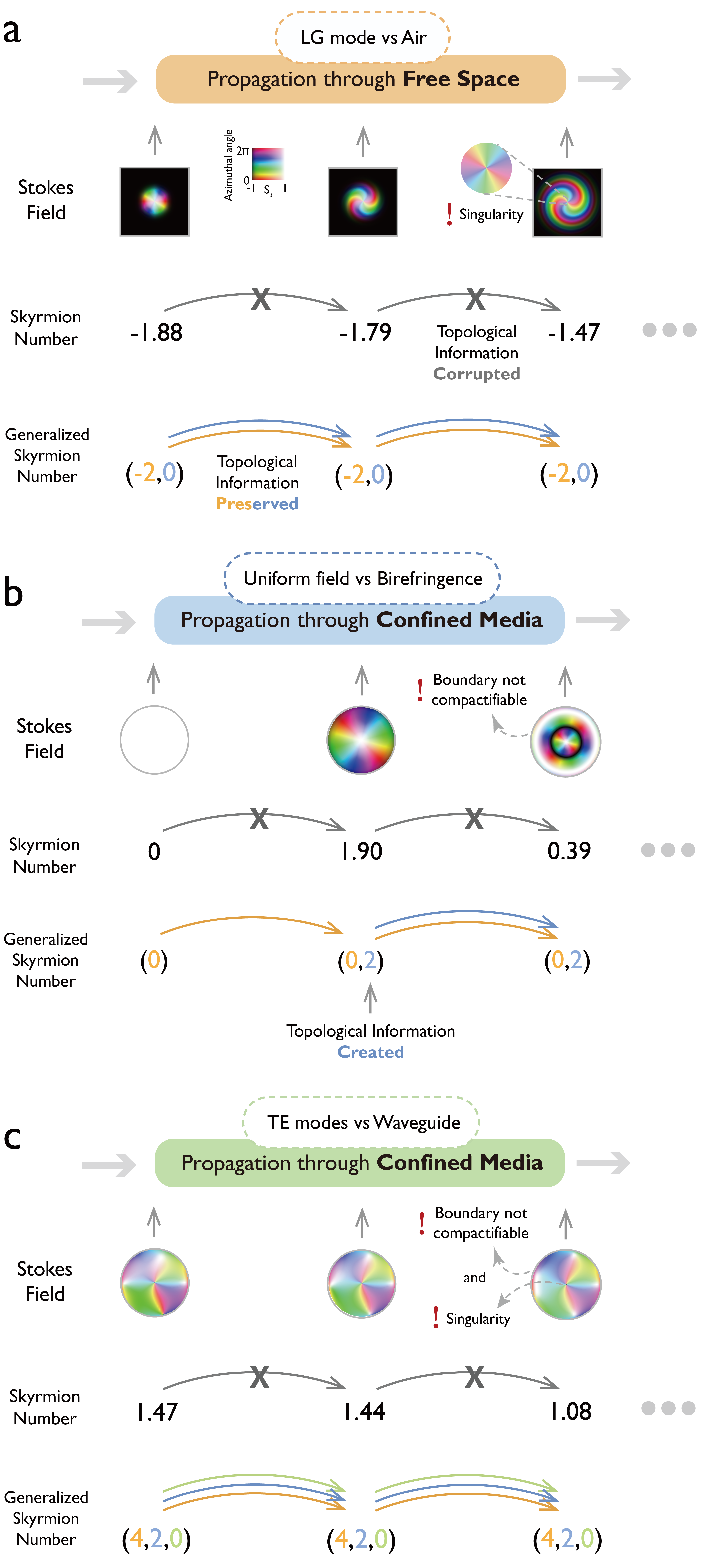}
\end{figure}
\captionof{figure}{\footnotesize {\bf Skyrmion numbers and generalized Skyrmion numbers of various polarization fields in propagation.} The transverse polarization profile of electromagnetic fields propagating through various media and their corresponding Skyrmion and generalized Skyrmion numbers. In each case, the usual Skyrmion number is not preserved in propagation as the field is not compactifiable. This is due to (a) the existence of singularities, (b) the lack of appropriate symmetries of the boundary, (c) or both. However, the generalized Skyrmion number remains constant. Throughout this paper, Stokes fields are depicted using hue to signify azimuthal angle $\tan\theta = s_2/s_1$ and saturation to represent height $s_3$ (similar to \cite{shen_optical_2023}). {\bf a}, Free space propagation of a superposition of two LG modes, $(p=0,l=-1)$ and $(p=0,l=1)$ with different beam widths. {\bf b}, Uniform right circularly polarized light passing through a spatially varying birefringent medium. In this case, the medium creates a new topological number with a charge of 2, but notice that the Skyrmion number of the incident beam is not lost. This demonstrates both the robustness of the generalized Skyrmion number and its ability to record topological information created by a medium. {\bf c}, A superposition of $\text{TE}_{21}$ and $\text{TE}_{31}$ modes in a cylindrical conducting waveguide. 
\label{fig: Concept1}}
\end{center}

Before proceeding to a more detailed description of the generalized Skyrmion number, observe in Fig.\ \ref{fig: Concept1} various physically significant real-world situations where the usual Skyrmion number is not preserved, but our newly defined topological number is. Heuristically, the stability of the generalized Skyrmion number in these examples is a consequence of its integer-valued nature in the presence of singularities (Fig.\ \ref{fig: Concept1}a, c) and inadequate boundary conditions (Fig.\ \ref{fig: Concept1}b, c), which are the two key situations that impede compactification. 

\begin{figure}[!ht]
\centering
\includegraphics[width=0.5\textwidth]{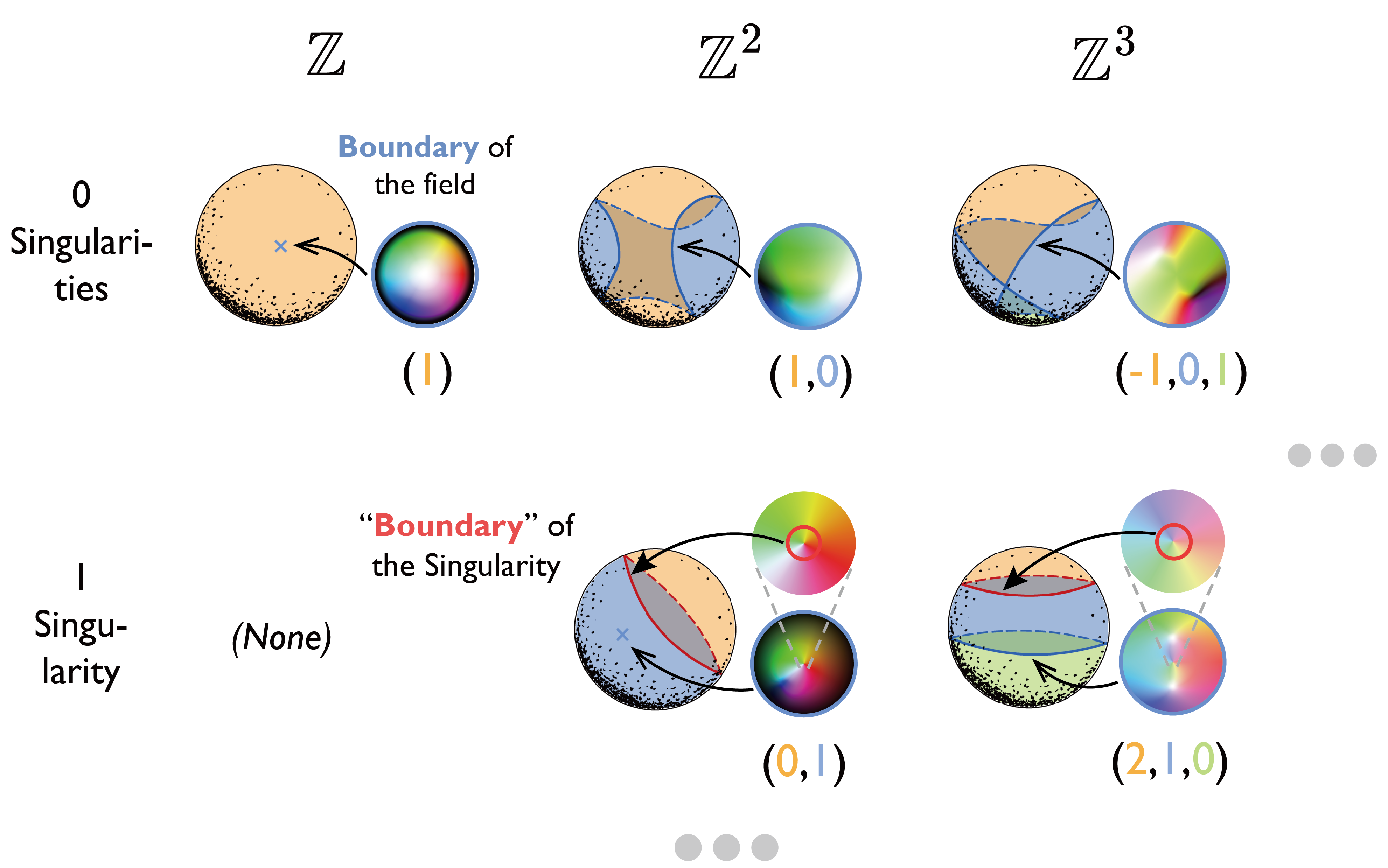}
\caption{\footnotesize {\bf Definition of the generalized Skyrmion number.} Various different partitions of a sphere that can be carved out by the boundary of an $S^2$-valued field. For each connected component of the partition, one topological number can be defined. Singularities can be regarded as ``boundaries'' by considering a small loop centered at the singularity. The top row shows scenarios without singularities, and the bottom row scenarios with a single singularity. Example Stokes fields realizing these partitions and their corresponding generalized Skyrmion numbers are also shown. 
\label{fig: Concept2}}
\end{figure}

We now provide a qualitative description of the generalized Skyrmion number, with mathematical details deferred to Methods 2. Given a field taking values in a compact, connected, orientable, smooth, $n$-dimensional manifold $Y$, consider the image traced out by the boundary of the field in $Y$ (See Fig.\ \ref{fig: Concept2} for the case $Y=S^2$). This partitions $Y$ into several different connected components. We can then assign, to each connected component, a topological number given by the number of times the field ``wraps'' around this specific component.

As with the usual Skyrmion number, we can retrieve this topological number through an integral equation by restricting equation (\ref{eq:Skyrmion Number}) to forms $\omega\in\Lambda^n(Y)$ compactly supported on our component of choice. For example, for $S^2$-valued fields, take any smooth $f\colon S^2 \longrightarrow \mathbb{R}$ with appropriate support. Then $\omega = f\omega_0$, where $\omega_0$ is the standard normalized volume form on $S^2$ arising from its usual Riemannian metric, defines the integer-valued integral equation
\begin{equation}
    \frac{1}{c}\int_{\Omega}f(\mathcal{S})\mathcal{S}\cdot\left(\frac{\partial \mathcal{S}}{\partial x}\times\frac{\partial \mathcal{S}}{\partial y}\right) dx dy,
\end{equation}
where $c = \int_0^{2\pi}\int_0^{\pi} f(\theta, \phi)\sin\theta d\theta d\phi$. Importantly, this defines $k$ different independent topological numbers, one for each connected component of $Y$ carved out by the boundary of the field (Fig.\ \ref{fig: Concept2}). Note that for fields constant on their boundary, the generalized Skyrmion number agrees with the usual Skyrmion number defined through compactification.

We can extend our considerations to fields with singularities by considering the image traced out by an infinitesimally small ball around the singularity. In this way, one can regard a singularity as simply another boundary (Fig.\ \ref{fig: Concept2}). 

To understand the topological robustness of the generalized Skyrmion number in optical polarization fields, we turn to the smoothness of the electric field in propagation brought on by the ellipticity of the Helmholtz equation (in the sense of PDE theory). This smoothness ensures that the curve traced out on the Poincar\'{e} sphere also varies smoothly in propagation \cite{evans}. Consequently, by considering the transformation of a single connected component, there must certainly exist a region of overlap contained within every transformation of this component, at least for some finite distance (Extended Fig.\ \ref{fig: Concept4}). By selecting a form with compact support in this overlap, one then has, via equation (\ref{eq:Skyrmion Number}), an integral equation that is guaranteed to be integer-valued for each transverse plane in propagation, precisely what we were looking for (see Methods 2, Corollary 4 for mathematical details). This, therefore, establishes topological protection of the generalized Skyrmion number associated with that connected component. By breaking propagation over long distances into smaller intervals where the above argument holds, one has that as long as the number of connected components carved out by the boundaries remains unchanged, topological protection holds. 

Through the discussion above, it is also easy to understand when topological protection does not hold. In particular, topological information can be created if the number of connected components increases, such as through the self-intersection of a curve, and destroyed if the number of connected components decreases, such as if a curve collapses into a point. Importantly, such phenomena are situations where the qualitative behavior of the curves traced by the boundary changes and are genuine topological phase transitions (See Extended Fig.\ \ref{fig: Concept3}).

As a proof of concept, we present experimental evidence of the robustness of the generalized Skyrmion number in free space propagation. In our experiments, we generated complex structured beams through spatially varying retarders, namely a gradient index lens that produces fields that cannot be compactified \cite{Chao2019} and a vortex half-wave plate that introduces a singularity \cite{rong_super-resolution_2015}. 

The generated beam (633nm) has an initial width of 7.2mm, and the transverse polarization profile is measured using traditional Stokes polarimetry \cite{he_polarisation_2021} at regular intervals between 0 and 2m. From these polarimetric measurements, both the usual Skyrmion number and the generalized Skyrmion number are extracted via the procedure described in Methods 3. Note that we limited our measurements to 2m because of constraints imposed by the size of our optical table, however, since the transverse polarization profile of the beam no longer changes significantly in the far field, we heuristically expect topological protection to persist for longer distances. We leave investigations into the limits of topological protection in propagation as future work.

Our results are given in Fig.\ \ref{fig: Experiment}, which includes both a normalized intensity distribution, the measured Stokes fields, and the computed Skyrmion numbers at various distances. The beams in experiments A and B were generated with a gradient index lens and incident right circularly polarized and horizontally polarized light, respectively. In both cases, the output field is not compactifiable, and the usual Skyrmion number (indicated by the solid line) in experiment A is seen to decrease with distance. 

Curiously, the usual Skyrmion number in experiment B does not seem to fluctuate as much. This is likely due to a symmetry property of the field, where the Skyrmion density of the upper left and lower right quadrants coincidentally cancels with the Skyrmion density of the upper right and lower left quadrants of the field. In either case, the computed generalized Skyrmion numbers remain far more stable in propagation, demonstrating the ability of the generalized Skyrmion number to overcome non-compactifiability. 

\begin{center}
\begin{figure}[!ht]
\includegraphics[width=0.5\textwidth]{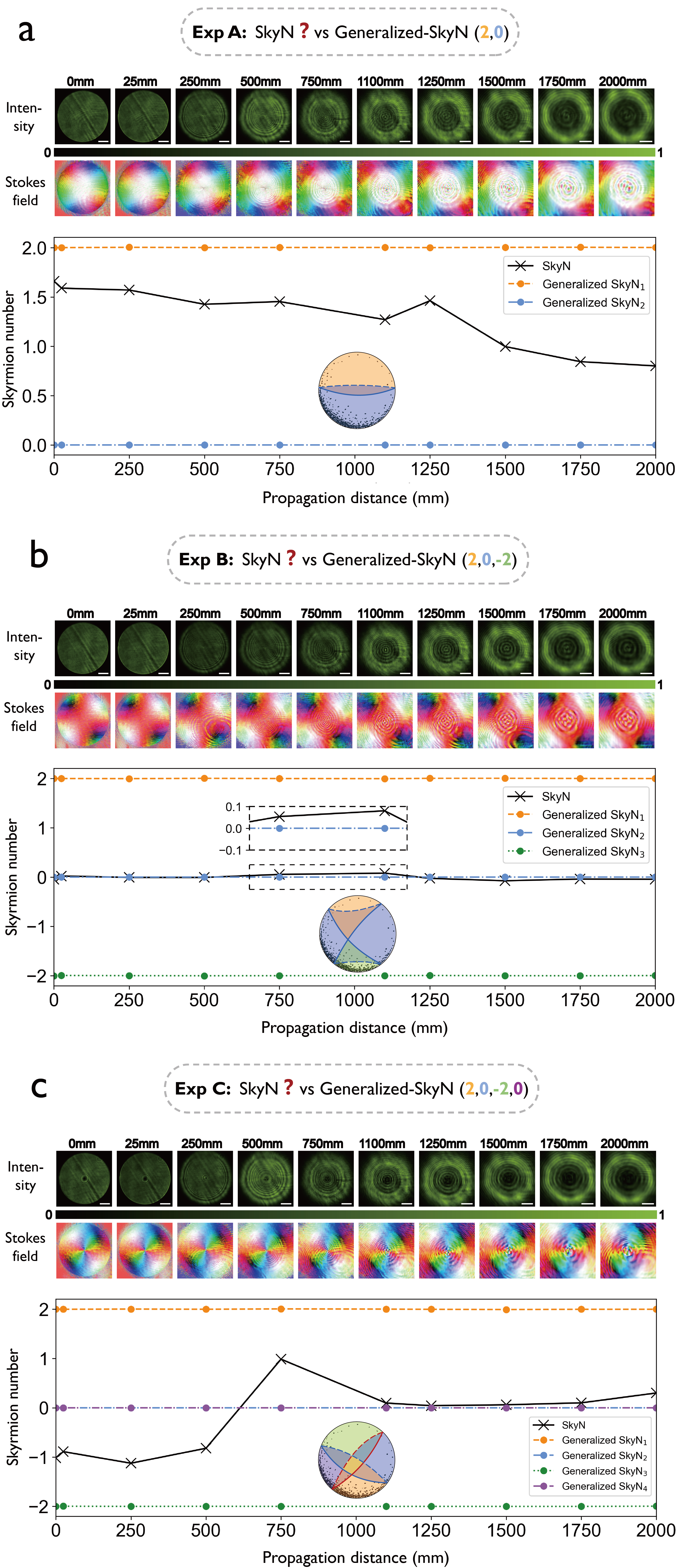}
\end{figure}
\captionof{figure}{\footnotesize {\bf Experimentally measured fields of three different vector beams in propagation and their respective Skyrmion and generalized Skyrmion numbers.} The Skyrmion number and generalized Skyrmion number computed from polarimetric measurements for three different vector beams in free space propagation at regular intervals between 0 and 2m. Normalized intensities and measured Stokes fields are also shown. Lastly, a depiction of the partitions of the Poincar\'{e} sphere associated with each generalized Skyrmion number is presented, with colors on the sphere corresponding to those used in the plots. {\bf a}, A non-compactifiable vector beam generated with uniform right circularly polarized light through a gradient index lens. Notice that the usual Skyrmion number decreases with distance, but the generalized Skyrmion number remains stable at $(2,0)$ for all distances. {\bf b}, A non-compactifiable vector beam generated with uniform horizontally polarized light through a gradient index lens. In this case, the usual Skyrmion number is coincidentally stabilized by a symmetry of the field, as explained in the main text. However, the generalized Skyrmion number still outperforms the usual Skyrmion number in terms of stability as highlighted by the inset, remaining close to $(2,0,-2)$ for all distances. Additionally, with three topological numbers compared to just one, the generalized Skyrmion offers greater advantages in terms of data density. {\bf c}, A non-compactifiable vector beam possessing a polarization singularity generated with a gradient index lens placed before a vortex half-wave plate and incident uniform horizontally polarized light. In this case, the usual Skyrmion number fluctuates with no clear trends, while the generalized Skyrmion number remains stable at $(2,0,-2,0)$ for all distances. 
\label{fig: Experiment}}
\end{center} 

The beam in experiment C was generated with a gradient index lens placed before a vortex half-wave plate with incident horizontally polarized light. The vortex plate introduces a singularity into the field, and the usual Skyrmion number is seen to fluctuate significantly in propagation. On the other hand, the generalized Skyrmion numbers remain very stable. This result demonstrates the ability of the generalized Skyrmion number to tolerate both singularities in the field as well as the non-compactifiability of its boundary. 

To conclude, this paper defines a new $\coprod_{i=1}^\infty \mathbb{Z}^i$-valued topological number for fields with a compact, connected, orientable, smooth, $n$-dimensional target manifold. When applied to optical polarization fields, our theory addresses one of the central problems concerning optical Skyrmions, namely the preservation of its topological character in propagation. 

More specifically, (1) we developed tools to establish the topological robustness of the generalized Skyrmion number applicable to general settings including the propagation of optical polarization fields in complex media, and (2) discussed situations where topological information is created and destroyed, offering strategies for manipulating the generalized Skyrmion number that can be explored further. Lastly, (3) observe that complicated partitions containing many connected components can be constructed with just one or two boundaries. Since each connected component corresponds to an independent topological number, this implies that introducing singularities into a field is a practical way to dramatically increase the topological information stored within it at the cost of a minimal increase in the field's complexity.

Combining this immense potential for representing high-density data with a theory of topological protection, creation, and destruction, the work presented in this paper represents a promising first step into unlocking the real-world use of optical Skyrmions in applications ranging from communications to computing.

Lastly, we stress that our definition of the generalized Skyrmion number is completely abstract and can be applied across diverse domains, ranging from concrete vector fields, such as velocity fields, to more abstract fields, such as fields of material parameters. The theory we have presented is, therefore, a broad mathematical theory that allows one to assign arbitrary fields with topological character and to analyze these fields by borrowing tools and techniques from topology and differential geometry. In this sense, our work aligns more closely with the principles of index theory within dynamical systems, the study of which has led to many deep results on the behavior of real-world systems across engineering, physics, chemistry, biology, and economics \cite{strogatz2015}. Notable examples include the esteemed Poincar\'{e}-Bendixson theorem, which uses topology to characterize the long-term behavior of dynamical systems on the plane. Much like index theory, which can be applied independently of the underlying physics, so too can our theory of generalized Skyrmions. In this paper, we have merely chosen optical polarization fields as one of many physically significant phenomena to study, and as an expository tool to demonstrate the power of the generalized Skyrmion number. 

\section*{Acknowledgement}
The authors would like to thank the support of St John’s College, the University of Oxford, and the Royal Society (R1 241734) (C.H.); Shenzhen Key Fundamental Research Project (No. JCYJ20210324120012035) (H.H.). Guangdong Major Project of Basic and Applied Basic Research (No. 2020B030103000) and Shenzhen University 2035 Initiative (2023B004) (Z.X., P.S. and X.Y.).

\bibliographystyle{unsrt}
{\footnotesize \bibliography{main}}

\begin{thebibliography}{10}

\bibitem{al_khawaja_skyrmions_2001}
U.~Al~Khawaja and H.~Stoof.
\newblock Skyrmions in a ferromagnetic {Bose}–{Einstein} condensate.
\newblock {\em Nature}, 411(6840):918--920, 2001.

\bibitem{Fert2017}
A.~Fert, N.~Reyren, and V.~Cros.
\newblock Magnetic skyrmions: Advances in physics and potential applications.
\newblock {\em Nature Reviews Materials}, 2(7):17031, 2017.

\bibitem{durrer_cosmic_2002}
R.~Durrer, M.~Kunz, and A.~Melchiorri.
\newblock Cosmic structure formation with topological defects.
\newblock {\em Physics Reports}, 364(1):1--81, 2002.

\bibitem{weintraub_differential_2014}
S.~H. Weintraub.
\newblock {\em Differential {Forms}: {Theory} and {Practice}}.
\newblock Academic Press, 2nd edition, 2014.

\bibitem{strogatz2015}
S.~H. Strogatz.
\newblock {\em {Nonlinear} {Dynamics} and {Chaos}: {With} {Applications} to {Physics}, {Biology}, {Chemistry}, and {Engineering}}.
\newblock CRC Press, 2nd edition, 2015.

\bibitem{donaldson1983application}
S.~K. Donaldson.
\newblock An application of gauge theory to four-dimensional topology.
\newblock {\em Journal of Differential Geometry}, 18(2):279--315, 1983.

\bibitem{skyrme1961non}
T.~H.~R. Skyrme.
\newblock A non-linear field theory.
\newblock {\em Proceedings of the Royal Society of London. Series A. Mathematical and Physical Sciences}, 260(1300):127--138, 1961.

\bibitem{Nagaosa2013}
N.~Nagaosa and Y.~Tokura.
\newblock Topological properties and dynamics of magnetic skyrmions.
\newblock {\em Nature Nanotechnology}, 8(12):899--911, 2013.

\bibitem{Tsesses2018}
S.~Tsesses et~al.
\newblock Optical skyrmion lattice in evanescent electromagnetic fields.
\newblock {\em Science}, 361(6406):993–--996, 2018.

\bibitem{shen_optical_2023}
Y.~Shen et~al.
\newblock Optical skyrmions and other topological quasiparticles of light.
\newblock {\em Nature Photonics}, 18(1):15--25, 2023.

\bibitem{He2022}
C.~He, Y.~Shen, and A.~Forbes.
\newblock Towards higher-dimensional structured light.
\newblock {\em Light: Science \& Applications}, 11(1):205, 2022.

\bibitem{Shen2021}
Y.~Shen.
\newblock Topological bimeronic beams.
\newblock {\em Opt. Lett.}, 46(15):3737--3740, 2021.

\bibitem{Shen2021_Super}
Y.~Shen et~al.
\newblock Supertoroidal light pulses as electromagnetic skyrmions propagating in free space.
\newblock {\em Nature Communications}, 12(1):5891, 2021.

\bibitem{Shen2023}
Y.~Shen et~al.
\newblock Topologically controlled multiskyrmions in photonic gradient-index lenses.
\newblock {\em Physical Review Applied}, 21(2):024025, 2024.

\bibitem{Bai2020}
C.~Bai et~al.
\newblock Dynamic tailoring of an optical skyrmion lattice in surface plasmon polaritons.
\newblock {\em Optics Express}, 28(7):10320, 2020.

\bibitem{Lin2021}
W.~Lin et~al.
\newblock Microcavity-based generation of full poincar\'e beams with arbitrary skyrmion numbers.
\newblock {\em Physical Review Research}, 3(2):023055, 2021.

\bibitem{he2023universal}
C.~He et~al.
\newblock A universal optical modulator for synthetic topologically tuneable structured matter.
\newblock Preprint at \url{https://arxiv.org/abs/2311.18148}, 2023.

\bibitem{du_deep-subwavelength_2019}
L.~Du et~al.
\newblock Deep-subwavelength features of photonic skyrmions in a confined electromagnetic field with orbital angular momentum.
\newblock {\em Nature Physics}, 15(7):650--654, 2019.

\bibitem{lei_photonic_2021}
X.~Lei et~al.
\newblock Photonic {Spin} {Lattices}: {Symmetry} {Constraints} for {Skyrmion} and {Meron} {Topologies}.
\newblock {\em Physical Review Letters}, 127(23):237403, 2021.

\bibitem{shi_spin_2021}
P.~Shi, L.~Du, and X.~Yuan.
\newblock Spin photonics: from transverse spin to photonic skyrmions.
\newblock {\em Nanophotonics}, 10(16):3927--3943, 2021.

\bibitem{teng_physical_2023}
H.~Teng et~al.
\newblock Physical conversion and superposition of optical skyrmion topologies.
\newblock {\em Photonics Research}, 11(12):2042--2053, 2023.

\bibitem{wang2024unlock}
A.~A. Wang et~al.
\newblock Unlocking new dimensions in photonic computing using optical skyrmions.
\newblock Preprint at \url{https://arxiv.org/abs/2407.16311}, 2024.

\bibitem{Cisowski2023}
C.~Cisowski, C.~Ross, and S.~Franke-Arnold.
\newblock Building paraxial optical skyrmions using rational maps.
\newblock {\em Advanced Photonics Research}, 4(4):2200350, 2023.

\bibitem{mcwilliam_topological_2023}
A.~McWilliam et~al.
\newblock Topological {Approach} of {Characterizing} {Optical} {Skyrmions} and {Multi}-{Skyrmions}.
\newblock {\em Laser \& Photonics Reviews}, 17(9):2300155, 2023.

\bibitem{shi_strong_2020}
P.~Shi, L.~Du, and X.~Yuan.
\newblock Strong spin–orbit interaction of photonic skyrmions at the general optical interface.
\newblock {\em Nanophotonics}, 9(15):4619--4628, 2020.

\bibitem{Gao2020}
S.~Gao et~al.
\newblock Paraxial skyrmionic beams.
\newblock {\em Physical Review A}, 102(5):053513, 2020.

\bibitem{ye2024theory}
Z.~Ye et~al.
\newblock Theory of paraxial optical skyrmions.
\newblock Preprint at \url{https://arxiv.org/abs/2404.11530}, 2024.

\bibitem{Liu2022}
C.~Liu et~al.
\newblock Disorder-induced topological state transition in the optical skyrmion family.
\newblock {\em Physical Review Letters}, 129(26):267401, 2022.

\bibitem{PhysRevLett.132.054003}
D.~A. Smirnova, F.~Nori, and K.~Y. Bliokh.
\newblock Water-wave vortices and skyrmions.
\newblock {\em Physical Review Letters}, 132(5):054003, 2024.

\bibitem{foster_two-dimensional_2019}
D~Foster et~al.
\newblock Two-dimensional skyrmion bags in liquid crystals and ferromagnets.
\newblock {\em Nature Physics}, 15(7):655--659, 2019.

\bibitem{acoustic}
H.~Ge et~al.
\newblock Observation of {Acoustic} {Skyrmions}.
\newblock {\em Physical Review Letters}, 127(14):144502, 2021.

\bibitem{Hurtado_acoustic}
R.~D. Muelas-Hurtado et~al.
\newblock {Observation} of {Polarization} {Singularities} and {Topological} {Textures} in {Sound} {Waves}.
\newblock {\em Physical Review Letters}, 129(20):204301, 2022.

\bibitem{Cao_Skyrmion}
L.~Cao et~al.
\newblock Observation of phononic skyrmions based on hybrid spin of elastic waves.
\newblock {\em Science Advances}, 9(7):eadf3652, 2023.

\bibitem{yang_scalar_2023}
B.~Yang et~al.
\newblock Scalar topological photonic nested meta-crystals and skyrmion surface states in the light cone continuum.
\newblock {\em Nature Materials}, 22(10):1203--1209, 2023.

\bibitem{ornelas_non-local_2024}
P.~Ornelas et~al.
\newblock Non-local skyrmions as topologically resilient quantum entangled states of light.
\newblock {\em Nature Photonics}, 18(3):258--266, 2024.

\bibitem{naber_topology_2011}
G.~L. Naber.
\newblock {\em Topology, {Geometry} and {Gauge} fields {Interactions}}.
\newblock Applied {Mathematical} {Sciences}. Springer, 2nd edition, 2011.

\bibitem{wang2024topological}
A.~A. Wang et~al.
\newblock Topological protection of optical skyrmions through complex media.
\newblock Preprint at \url{https://arxiv.org/abs/2403.07837}, 2023.

\bibitem{shi2024embarking}
L.~Shi, Z.~Che, and Y.~Kivshar.
\newblock Embarking on a skyrmion odyssey.
\newblock {\em Photonics Insights}, 3(1):C02--C02, 2024.

\bibitem{evans}
L.~C. Evans.
\newblock {\em Partial Differential Equations}, volume~19 of {\em Graduate Studies in Mathematics}.
\newblock American Mathematical Society, 2nd edition, 2010.

\bibitem{Chao2019}
C.~He et~al.
\newblock Complex vectorial optics through gradient index lens cascades.
\newblock {\em Nature Communications}, 10(1):4264, 2019.

\bibitem{rong_super-resolution_2015}
Z.~Rong et~al.
\newblock Super-resolution microscopy based on fluorescence emission difference of cylindrical vector beams.
\newblock {\em Optics Communications}, 354:71--78, 2015.

\bibitem{he_polarisation_2021}
C.~He et~al.
\newblock Polarisation optics for biomedical and clinical applications: a review.
\newblock {\em Light: Science \& Applications}, 10(1):194, 2021.

\end{thebibliography}


\begin{thebibliography}{1}

\bibitem{hatcher_algebraic_2001}
A.~Hatcher.
\newblock {\em Algebraic {Topology}}.
\newblock Cambridge University Press, 2001.

\bibitem{Coope}
I.~D. Coope.
\newblock Circle fitting by linear and nonlinear least squares.
\newblock {\em J. Optim. Theory Appl.}, 76(2):381--388, 1993.

\end{thebibliography}

\section*{Methods}
\setcounter{section}{0}

\section{The Topological Protection of Optical Skyrmions}
To date, investigations into the topological stability of the optical Skyrmion in propagation are still in their infancy. Nonetheless, two major competing factors have emerged, one in favor of topological robustness and the other against it. The argument centers around the homotopy invariance of the degree \citesupp{hatcher_algebraic_2001}, which is the mathematical abstraction that quantifies topological protection against deformations of the field. 

More intuitively, this argument can be understood in the following way. Suppose one can guarantee that equation (\ref{eq:Skyrmion Number}) evaluates to an integer for some specific normalized $\omega \in \Lambda^n(Y)$. Then, as the integrand in this equation varies smoothly with respect to $\mathcal{S}$, the assignment of the field to its Skyrmion number must also, in somewhat imprecise language, vary smoothly with $\mathcal{S}$. However, as the integral equation evaluates to an integer, it must be that smoothly deforming $\mathcal{S}$ cannot change its Skyrmion number, as it is impossible to smoothly transition from one integer to another without also passing through every number in between.

Applying the above argument to optical polarization fields in propagation, it is clear that the underlying Helmholtz equation solves half of the puzzle. This is because the ellipticity of the Helmholtz equation guarantees the smoothness of the electric field in propagation, which provided there are no zeros of the electric field, descends onto a smooth homotopy of polarization fields via the Hopf map. The other half of the puzzle requires the existence of a normalized form $\omega \in \Lambda^2(S^2)$ for which equation (\ref{eq:Skyrmion Number}) evaluates to an integer on every transverse plane. Consequently, if the transverse polarization profile remains compactifiable in propagation, as is the case for important classes of vector beams, then the argument above establishes topological protection (see Methods 2). 

Here, another distinguishing property of the optical Skyrmion rears its ugly head: unlike more traditional Skyrmions, optical Skyrmions are not topological solitons in a true sense and, therefore, are not guaranteed to remain compactifiable in propagation without intentional design. In particular, the existence of polarization singularities and a loss of symmetry of the field in propagation impede compactifiability in all but a narrow number of situations. Note also that this peculiarity of the optical Skyrmion is responsible for the heavily mathematical approach introduced in this work, as more physical notions of energy stability do not naturally apply. 

However, while this lack of energy stability has implications for the topological robustness of optical Skyrmions, it also provides an unexpected advantage in terms of data density, where energy constraints do not impede the formation of diverse and complex optical Skyrmions of high orders. Together with the increasingly flexible techniques for beam generation, this suggests that overcoming the need for compactification in establishing the preservation of the Skyrmion number holds significant potential to transform optical Skyrmions from an interesting theoretical phenomenon to real high-density data applications. 

\section{Defining the Generalized Skyrmion Number}

In this admittedly long and technical section, we build up to the definition of the generalized Skyrmion number in three steps. To facilitate our discussion, we begin by providing a brief overview of our approach, including key terminology and notation. In this section, we consider smooth fields $\mathcal{S}\colon\Omega \longrightarrow Y$ where $\Omega$ is an open subset of $\mathbb{R}^n$ with the usual orientation and $Y$ a compact, connected, oriented, smooth, $n$-dimensional manifold. When specializing our theory to optical polarization fields, we will explicitly state so, and make clear that, in fact, $Y=S^2$. Lastly, we denote the open ball of radius $r$ centered about $p\in \mathbb{R}^n$ by $B_r(p)\subset \mathbb{R}^n$. 

Our three main steps are as follows: Firstly, we generalize the compactification process introduced in \cite{wang2024topological} to fields taking values in an arbitrary manifold $Y$ and define the notion of generalized $Y$-valued Skyrmions for which equation (\ref{eq:Skyrmion Number}) is guaranteed to be integer-valued.

Secondly, we consider the case where fields are no longer compactifiable, and for each normalized $\omega\in \Lambda^n(Y)$, assign to the possibly non-compactifiable field $\mathcal{S}$ a corresponding $\omega$-Skyrmion number. Next, by replacing compactifiability with the more general notion of $(\psi,F)$-extendability, we recover some way of interpreting the $\omega$-Skyrmion number topologically.

Thirdly, by examining the qualitative behavior of the $\omega$-Skyrmion number as $\omega$ varies across $\Lambda^n(Y)$, we show that for each $\mathcal{S}$, it is possible to partition $\Lambda^n(Y)$ into collections of forms for which the corresponding notion of $\omega$-Skyrmion number is consistent and integer-valued. Each partition, which can be associated with a connected component carved out by the boundary of the field, then defines a single topological number. We then formalize this observation by turning to the De Rham cohomology of compactly supported forms \cite{weintraub_differential_2014}. Lastly, we provide a method of establishing the topological robustness of the generalized Skyrmion number against deformations of the field using appropriate $(\psi, F)$-extensions. 

In each of the three steps, we prove an abstract result of topological protection relevant to the generality of the Skyrmion number defined in that section, and highlight these results as {\bf Theorems}. We then ground our abstract results in reality by providing a description of how they can be applied to optical polarization fields.

Lastly, in our exposition of the topic, we will, at times, be loose with distinctions between smoothness and continuity, noting that appropriate cohomology theories can stand in for the De Rham cohomology to make everything rigorous. 

\subsection{\texorpdfstring{$Y$}{Y}-Valued Skyrmions}

Let $\mathcal{S}\colon \Omega \longrightarrow Y$ be a smooth map. Suppose there exists a compact, connected, oriented, smooth $n$-dimensional manifold $X$ and a smooth map $\psi \colon \Omega \longrightarrow X$ such that 
\begin{enumerate}
    \item $\psi$ is an orientation preserving diffeomorphism of $\Omega$ onto its image,
    \item $\psi(\Omega)$ is a dense subset of $X$ with full measure,
    \item $\mathcal{S}\circ \psi^{-1}$ extends via continuity to a smooth map $\tilde{\mathcal{S}}$ on all of $X$,
\end{enumerate}
then we say $\mathcal{S}$ is $\psi$-compactifiable, and call $\tilde{\mathcal{S}}$ the $\psi$-compactification of $\mathcal{S}$. If $\mathcal{S}$ is $\psi$-compactifiable, we define its Skyrmion number by
\begin{equation}
    \deg \mathcal{S} \coloneqq \deg \tilde{\mathcal{S}},
\end{equation}
where the $\deg$ on the right-hand side of the equation is the usual degree arising from De Rham cohomology \cite{naber_topology_2011}. The topological robustness of the Skyrmion then arises from the homotopy invariance of the degree, where given fields $\mathcal{S}_1$ and $\mathcal{S}_2$ which are $\psi_1$- and $\psi_2$-compactifiable, respectively, with $\operatorname{codom}\psi_1 = \operatorname{codom}\psi_2$, then $\deg \mathcal{S}_1 = \deg \mathcal{S}_2$ whenever $\tilde{\mathcal{S}}_1$ and $\tilde{\mathcal{S}}_2$ are smoothly homotopic. If, further, $X=Y=S^n$, we have, via the Hopf degree theorem, the converse statement that whenever $\deg \mathcal{S}_1 = \deg \mathcal{S}_2$, $\tilde{\mathcal{S}}_1$ and $\tilde{\mathcal{S}}_2$ are smoothly homotopic. With the machinery above, we trivially have the following result.\\

\noindent {\bf Theorem 1.} Let $\Omega_z$ be an open subset of $\mathbb{R}^n$ for each $z\in[z_0,z_f]$ and $\mathcal{S}^z\colon \Omega_z \longrightarrow Y$ a smooth map. Suppose, for each $z \in [z_0, z_f]$, $
\mathcal{S}^z$ is compactifiable via $\psi_z\colon \Omega_z \longrightarrow X$. If further $H\colon X \times [z_0, z_f] \longrightarrow Y$, $H(x, z) = \tilde{\mathcal{S}}^z(x)$ is smooth, then $\deg \mathcal{S}^{z_1} = \deg \mathcal{S}^{z_2}$ for all $z_1, z_2 \in [z_0, z_f]$.

Specializing to optical polarization fields, the theorem above makes rigorous the claim in Methods 1 of topological protection for fields whose transverse polarization profile remains compactifiable in propagation. In the case of vector beams formed from a superposition of LG modes, the polarization state at infinity depends on the asymptotic behavior of the individual modes. If, for example, a single mode dominates in the far field, and provided there are no polarization singularities, taking $\Omega_z = \mathbb{R}^2$ and $\psi_z$ to be any inverse stereographic projection establishes topological protection by the theorem presented above.  

\subsection{The \texorpdfstring{$\omega$}{w}-Skyrmion Number and \texorpdfstring{$(\psi, F)$}{(p,f)}-Extensions}

Without compactifiability, equation (\ref{eq:Skyrmion Number}) is no longer guaranteed to be the same for all normalized $\omega \in \Lambda^n(Y)$. Nevertheless, we can still define, for each normalized $\omega \in \Lambda^n(Y)$, a form-dependent Skyrmion number 
\begin{equation}
    \operatorname{SkyN}_\omega(\mathcal{S}) \coloneqq \int_\Omega \mathcal{S}^\ast \omega,
\end{equation}
which we call the $\omega$-Skyrmion number of $\mathcal{S}$. To interpret $\omega$-Skyrmion numbers topologically, we introduce the following extension process. Let $X$ be a compact, connected, oriented, smooth $n$-dimensional manifold and $\psi \colon \Omega \longrightarrow X$ be an orientation preserving diffeomorphism onto its image. We say $\mathcal{S}$ is $\psi$-extendable if 
\begin{enumerate}
    \item $\mathcal{S} \circ \psi^{-1}$ extends via continuity to $\overline{\psi(\Omega)}$ and
    \item there exists a smooth $F\colon X-\psi(\Omega) \longrightarrow Y$ such that $F\lvert_{\partial \psi(\Omega)} = \mathcal{S}\circ\psi^{-1}\lvert_{\partial \psi(\Omega)}$, where the latter is to be understood as the continuous extension of $\mathcal{S} \circ \psi^{-1}$ to $\overline{\psi(\Omega)}$.
\end{enumerate}
We then call
\begin{equation}
    \tilde{\mathcal{S}}_F = \left\{\begin{array}{r l}
        \mathcal{S} \circ \psi^{-1}, & x\in\overline{\psi(\Omega)} \\
        F, & x\in X-\psi(\Omega) 
    \end{array}\right.
\end{equation}
the $(\psi, F)$-extension of $\mathcal{S}$ to $X$, where $\tilde{\mathcal{S}}_F$ is continuous by a gluing argument and piecewise smooth by construction. 

Ignoring technicalities arising from piecewise smoothness, we can relate $\operatorname{SkyN}_\omega(\mathcal{S})$ to the degree of the map $\tilde{\mathcal{S}}_F$ via the equation 
\begin{equation}
    \deg \tilde{\mathcal{S}}_F = \operatorname{SkyN}_\omega(\mathcal{S}) + \int_{X-\psi(\Omega)} F^\ast \omega. \label{eq:omega-Skyrmion number}
\end{equation}
Notice further that whenever $\mathcal{S}$ is $\psi$-compactifiable, $\operatorname{SkyN}_\omega(\mathcal{S}) = \deg \mathcal{S}$ for all normalized $\omega \in \Lambda^n(Y)$ as the density of $\psi(\Omega)$ in $X$ implies that $\mathcal{S}$ extends to a unique $\tilde{\mathcal{S}}_F = \tilde{\mathcal{S}}$.

We now state our main topological robustness result for $\omega$-Skyrmion numbers.\\

\noindent {\bf Theorem 2.} Let $\Omega_z$ be an open subset of $\mathbb{R}^n$ for each $z\in[z_0,z_f]$ and $\mathcal{S}^z\colon \Omega_z \longrightarrow Y$ a smooth map. Suppose, for each $z\in[z_0,z_f]$, $\mathcal{S}^z$ is $\psi_z$-extendable, $\psi_z(\Omega_z)$ is the same for all $z$, and $J\colon \psi_z(\Omega_z)\times[z_0,z_f] \longrightarrow Y$, $J(x,z) = \mathcal{S}^z(\psi_z^{-1}(x))$ is smooth. If further $\mathcal{S}^z$ when extended to $\overline{\psi_z(\Omega_z)}$ restricts to the same function $\partial \mathcal{S}$ on $\partial\psi_z(\Omega_z)$ for all $z$, then $\operatorname{SkyN}_\omega(\mathcal{S}^{z_1}) = \operatorname{SkyN}_\omega(\mathcal{S}^{z_2})$ for all normalized $\omega \in \Lambda^n(Y)$, and $z_1, z_2 \in [z_0, z_f]$.\\

\noindent {\bf Proof:} Notice that any $F$ that extends $\mathcal{S}^z$ for some $z$ must necessarily also extend every other $\mathcal{S}^{z'}$, $z' \in [z_0, z_f]$. Picking one such $F$, set $H\colon X\times [z_0, z_f] \longrightarrow Y$ by $H(x,z) = \tilde{\mathcal{S}}^z_F(x)$. Notice that $H$ is a homotopy by the smoothness of $J$. The result then follows immediately from the homotopy invariance of the degree and equation (\ref{eq:omega-Skyrmion number}). $\blacksquare$

As a concrete application of the above to optical polarization fields, consider the situation of a cylindrical conducting waveguide of radius $a$ aligned parallel to the $z$-axis. Suppose we have a monochromatic wave of the form 
\begin{align}
    \bm{E}(x,y,z,t) &= \bm{E}_0(x,y,z)e^{-i\omega t} \\
    \bm{B}(x,y,z,t) &= \bm{B}_0(x,y,z)e^{-i\omega t}
\end{align}
confined within the conductor. The conductor then imposes the following boundary conditions: the parallel component of the electric field at the air-conductor interface $\bm{E}_{\parallel} = 0$, and the perpendicular component of the magnetic field at the air-conductor interface $B_\perp = 0$. Crucially, the boundary condition on the electric field implies that 
\begin{equation}
    \bm{E}_0(a\cos\theta,a\sin\theta, z) \propto \begin{pmatrix}
        \cos\theta \\ \sin \theta \\ 0
    \end{pmatrix}
\end{equation}
for any $z$. Suppose further that 
\begin{equation}
    \lVert E_x(x, y, z) + E_y(x, y, z) \rVert > 0
\end{equation}
for all $(x,y)\in \overline{B_a(0)}$ and $z\in[z_0, z_f]$, then, for each $z$, the transverse components of the electric field define for us a corresponding Stokes field $\mathcal{S}^z \colon \overline{B_a(0)}\longrightarrow S^2$ that satisfies 
\begin{equation}
\label{eq: waveguide boundary condition}
    \mathcal{S}^z(a\cos\theta, a\sin\theta) = \begin{pmatrix}
        \cos 2\theta \\ \sin 2\theta \\ 0
    \end{pmatrix}.
\end{equation}

To extend the field, let $\psi$ be the orientation preserving diffeomorphism 
\begin{equation}
    \psi(r,\theta) = \left(r\cos\theta, r\sin\theta, \sqrt{1-r^2}\right) \label{eq:psi}
\end{equation}
that maps $B_a(0)$ onto the open upper hemisphere $S^+ = \{(x^1, x^2, x^3) \in S^2\colon x^3 > 0\}$ of the Poincar\'{e} sphere. Taking $S^-=\{(x^1, x^2, x^3) \in S^2 \colon x^3<0\}$, we may construct a distinguished $F \colon S^- \longrightarrow S^2$ given by
\begin{multline}
    F\left(\sqrt{1-x_3^2}\cos\theta, \sqrt{1-x_3^2}\sin\theta, x_3\right) \\ = \left(\sqrt{1-x_3^2}\cos 2\theta, \sqrt{1-x_3^2}\sin 2\theta, x_3\right)
\end{multline}
that canonically extends every such $\mathcal{S}^z$. 

Setting $H\colon S^2 \times [z_0, z_f] \longrightarrow S^2$ as $H(x,z) = \tilde{\mathcal{S}}_F^{z}(x)$, we have a well-defined homotopy from $\tilde{\mathcal{S}}^{z_0}_F$ to $\tilde{\mathcal{S}}^{z_f}_F$. From $H$, we conclude that
\begin{equation}
    \operatorname{SkyN}_{\omega}(\mathcal{S}^{z_0}) = \operatorname{SkyN}_{\omega}(\mathcal{S}^{z_f})
\end{equation}
for all normalized $\omega \in \Lambda^2(S^2)$.

\subsection{The Generalized Skyrmion Number}

Building on the theory of $(\psi,F)$-extensions introduced in the previous section, we begin with the following observation. Suppose $\mathcal{S}$ has a $(\psi, F)$-extension where $F$ is not surjective. Then, any normalized $\omega \in \Lambda^n(Y)$ whose support $\operatorname{supp} \omega \subset Y - F(X-\psi(\Omega))$ satisfies the relation
\begin{equation}
    \deg \tilde{\mathcal{S}}_F = \operatorname{SkyN}_\omega(\mathcal{S}) + \int_{X-\psi(\Omega)} F^\ast \omega = \operatorname{SkyN}_\omega(\mathcal{S}).
\end{equation}
Therefore, for every $\omega$ with the stated restriction, the corresponding $\omega$-Skyrmion number is integer-valued. Moreover, any two forms sharing this restriction have the same $\omega$-Skyrmion number. This suggests that given a field $\mathcal{S}$, one way to define a generalized Skyrmion number is based on the largest suitable subset of forms for which the corresponding notion of $\omega$-Skyrmion number is consistent and integer-valued. 

To make this more precise, consider the following implication of the observation above. For ease of visualization, let $\mathcal{S}\colon B_1(0)\subset \mathbb{R}^2 \longrightarrow S^2$ and suppose $\mathcal{S}$ extends via continuity to $\overline{B_1(0)}$. Then $\partial \mathcal{S} \coloneqq \mathcal{S}\lvert_{\partial B_1(0)}$ traces out a curve in $S^2$ that partitions it into connected components $V_1, \ldots, V_k$. Since $\partial B_1(0)$ is compact, $\partial \mathcal{S}(\partial B_1(0))$ is also compact, and hence closed. This then implies that the $V_i$s are open sets. Let $p \in V_i$. By the openness of $V_i$, there is some $B_r(0)\subset \mathbb{R}^3$ so that $p \in B_r(0)\cap S^2 = U_i \subset V_i$. Consider now the following $(\psi,F)$-extension. Let $\psi$ be as defined in equation (\ref{eq:psi}) and $F\colon S^- \longrightarrow S^2$ be
\begin{multline}
    F\left(\sqrt{1-x_3^2}\cos\theta, \sqrt{1-x_3^2}\sin\theta, x_3\right) \\ = \frac{(1+x_3)\partial\mathcal{S}(\theta)+x_3p}{\lVert (1+x_3)\partial\mathcal{S}(\theta)+x_3p\rVert}. \label{eq: homotope}
\end{multline}
Then, by construction, $F(S^-)\cap U_i = \varnothing$. Therefore, for any normalized $\omega$ with compact support in $U_i$, $\operatorname{SkyN}_\omega(\mathcal{S})$ is consistent and integer-valued. Since the point $p$ was arbitrary, and $V_i$ is connected, one can essentially regard $\operatorname{SkyN}_\omega(\mathcal{S})$ as a topological number associated with $V_i$ rather than $\omega$. 

To formalize this intuition, we review some basic differential geometry. Let $X$ be an $n$-dimensional smooth manifold. We define by $\Lambda_c^k(X)$ the smooth real-valued $k$-forms on $X$ with compact support. Then the exterior differentiation maps $d^k \colon \Lambda_c^k(X) \longrightarrow \Lambda_c^{k+1}(X)$ form a cochain complex 
\begin{multline}
    \cdots \xrightarrow[]{} 0 \xrightarrow[]{} \Lambda_c^0(X) \xrightarrow[]{d^0} \Lambda_c^1(X) \xrightarrow[]{d^1} \cdots \\\xrightarrow[]{d^{n-1}} \Lambda_c^n(X) \xrightarrow[]{} 0 \xrightarrow[]{} \cdots
\end{multline}
from which we may define the de Rham cohomology groups with compact supports, $H_c^k(X) \coloneqq \ker d^k/\operatorname{im} d^{k+1}$ \cite{weintraub_differential_2014}. Let $f\colon X\longrightarrow Y$ be a smooth map. We call $f$ proper if for every compact set $L\subseteq Y$, $K=f^{-1}(L)$ is compact. Then, for each $k\in\mathbb{Z}$ and proper smooth map $f\colon X \longrightarrow Y$, we have a natural corresponding map in cohomology
\begin{equation}
    f^\#\colon H_c^k(Y) \longrightarrow H_c^k(X), \quad [\omega] \mapsto [f^\ast \omega].
\end{equation}
One may show that for any connected, oriented, smooth, $n$-dimensional manifold, $H_c^n(X) \cong \mathbb{R}$ via the isomorphism $\omega \mapsto \int_X\omega$. Therefore, any proper smooth map $f\colon X\longrightarrow Y$ between connected, oriented, smooth, $n$-dimensional manifolds induces a linear map 
\begin{equation}
    f^\# \colon H_c^n(Y) \cong \mathbb{R} \longrightarrow H_c^n(X) \cong \mathbb{R},
\end{equation}
which we may identify with a real number $c \in \mathbb{R}$ called the degree of $f$. As with the degree that originates from the De Rham cohomology, one can show that the degree originating from the De Rham cohomology of compactly support forms is also integer-valued. From this, we now have all the necessary ingredients to assemble the definition of the generalized Skyrmion number. 

As before, let $\Omega$ be an open subset of $\mathbb{R}^n$, $Y$ a compact, connected, oriented, smooth $n$-dimensional manifold, and $\mathcal{S}\colon \Omega \longrightarrow Y$ a smooth map. Let $\psi \colon \Omega \longrightarrow U \subset \mathbb{R}^n$ be an orientation preserving diffeomorphism onto a bounded subset $U$ of $\mathbb{R}^n$. Suppose that $\mathcal{S}\circ \psi^{-1}$ extends via continuity to $\partial U$ and denote by $\partial \mathcal{S}$ this continuous extension when restricted to $\partial U$. Then $\partial \mathcal{S}$ partitions $Y$ into open connected components $V_1, \ldots, V_k$ such that $\bigcup_{i=1}^k V_i = Y-\partial \mathcal{S}(\partial U)$. 

For each $i \in \{1,\ldots, k\}$, we may write $\mathcal{S}^{-1}(V_i) = \bigcup_{j=1}^{l_i} U_{ij}$ where $U_{i1}, \ldots U_{il_i}$ are connected components of $\mathcal{S}^{-1}(V_i)$. Clearly, $\mathcal{S}\lvert_{U_{ij}}$ is proper. We can therefore define the $i j$\textsuperscript{th} local Skyrmion number of $\mathcal{S}$ by 
\begin{equation}
    \operatorname{SkyN}_{ij}(\mathcal{S}) \coloneqq \deg \mathcal{S}\lvert_{U_{ij}},
\end{equation}
where we understand $\deg \mathcal{S}\lvert_{U_{ij}}$ as arising from the map  $\mathcal{S}\lvert_{U_{ij}}^\#\colon H_c^n(V_i) \longrightarrow H_c^n(U_{ij})$. We further define the $i$\textsuperscript{th} Skyrmion number of $\mathcal{S}$ by 
\begin{equation}
    \operatorname{SkyN}_i(\mathcal{S}) \coloneqq \left\{\begin{array}{r l}
        0, & \text{if $\mathcal{S}^{-1}(V_i)=\varnothing$}, \\
        \sum_{j=1}^{l_i} \operatorname{SkyN}_{ij}(\mathcal{S}), & \text{otherwise}. 
    \end{array} \right.
\end{equation}

Lastly, we define the generalized Skyrmion number of $\mathcal{S}$ by $\operatorname{SkyN}(\mathcal{S})\in \coprod_{i=1}^\infty \mathbb{Z}^i$, 
\begin{equation}
    \operatorname{SkyN}(\mathcal{S}) \coloneqq (\operatorname{SkyN}_1(\mathcal{S}), \ldots, \operatorname{SkyN}_{k}(\mathcal{S})),
\end{equation}
and call $\mathcal{S}$ a $Y$-valued generalized Skyrmion of class $k$.

We now state the primary topological robustness result for the generalized Skyrmion number. \\

\noindent {\bf Theorem 3.} Let $\Omega_z$ be an open subset of $\mathbb{R}^n$ for each $z\in[z_0,z_f]$ and $\mathcal{S}^z\colon \Omega_z\longrightarrow Y$ a smooth map. Let $V_i^z$ correspond to the connected component of $Y$ that defines the $i$\textsuperscript{th} generalized Skyrmion number of $\mathcal{S}^z$. Suppose there exists an open set $W\subset \bigcap_{z\in[z_0,z_f]}V_i^z$ and $(\psi_z, F_z)$-extensions of $\mathcal{S}^z$ so that 
\begin{enumerate}
    \item $F_z(X-\psi_z(\Omega_z)) \cap W = \varnothing$ for all $z \in [z_0, z_f]$ and
    \item $H\colon X\times [z_0, z_f]\longrightarrow Y$, $H(x,z) = \tilde{\mathcal{S}}^z_{F_z}(x)$ is smooth,
\end{enumerate}
then $\operatorname{SkyN}_i(\mathcal{S}^{z_1}) = \operatorname{SkyN}_i(\mathcal{S}^{z_2})$ for all $z_1, z_2 \in [z_0, z_f]$. \\

\noindent {\bf Proof:} Let $\omega \in \Lambda^n(Y)$ be normalized and compactly supported in $W$. Then we have the following chain of equalities
\begin{align}
    \operatorname{SkyN}_i(\mathcal{S}^{z_1}) \nonumber & = \operatorname{SkyN}_\omega(\mathcal{S}^{z_1}) \nonumber \\ & = \deg \tilde{\mathcal{S}}^{z_1}_{F_{z_1}} \nonumber \\ & =
    \deg \tilde{\mathcal{S}}^{z_2}_{F_{z_2}} \nonumber \\ & = \operatorname{SkyN}_\omega(\mathcal{S}^{z_2}) \nonumber \\ &= \operatorname{SkyN}_i(\mathcal{S}^{z_2}),
\end{align}
where $\deg \tilde{\mathcal{S}}^{z_1}_{F_{z_1}} = \deg \tilde{\mathcal{S}}^{z_2}_{F_{z_2}}$ comes from the homotopy invariance of the degree. $\blacksquare$

As a powerful corollary to this, when $Y=S^2$, we may use generalizations of the $(\psi, F)$-extension defined by equation (\ref{eq: homotope}) to prove the following.\\

\noindent {\bf Corollary 4.} Let $\Omega$ be a connected open subset of $\mathbb{R}^2$ and $\mathcal{S} \colon \Omega \times [z_0, z_f] \longrightarrow S^2$ a smooth map. Suppose $\psi\colon \Omega \longrightarrow U$ is an orientation preserving diffeomorphism onto a bounded subset $U$ of $\mathbb{R}^2$ so that $\mathcal{S}\lvert_{\Omega \times \{z\}}\circ \psi^{-1}$ extends via continuity to $\partial U$ for all $z\in[z_0, z_f]$. Let $V_i^z$ correspond to the connected component of $Y$ that defines the $i$\textsuperscript{th} generalized Skyrmion number of $\mathcal{S}^z\coloneqq \mathcal{S}\lvert_{\Omega\times \{z\}}$. Suppose there exists an open set $W\subset \bigcap_{z\in[z_0,z_f]}V_i^z$, then $\operatorname{SkyN}_i(\mathcal{S}^{z_1}) = \operatorname{SkyN}_i(\mathcal{S}^{z_2})$ for all $z_1, z_2 \in [z_0, z_f]$.\\

\noindent {\bf Proof:} Here, we give a somewhat qualitative proof. Notice first that $\partial U$ is homeomorphic to a disjoint union of circles $\partial U \cong \coprod_{j=1}^m S^1$. Let $\varphi$ be any inverse stereographic projection map. We construct a $(\varphi\circ\psi,F)$-extension in the following way. For each $j=1,\ldots,m$, pick a homeomorphism $\mathcal{C}_j\colon S^1 \longrightarrow \mathbb{R}^2$ that parameterizes the corresponding circle of $\partial U$. Then $\varphi \circ \mathcal{C}_j$ is a circle in $S^2$ which partitions $S^2$ into two components. By the connectedness of $\Omega$, $\varphi(\psi(\Omega))$ is contained in exactly one of these components. Now, construct a continuous homotopy $H_j\colon S^1\times[0,1]\longrightarrow S^2$ so that $H_j(s,0) = \varphi(\mathcal{C}_j(s))$, $H_j(s,1)$ is some constant $c_j$, $H_j\lvert_{S^1\times[0,1)}$ is injective, and that the image of $H_j$ is the union of the image of $\varphi\circ\mathcal{C}_j$ and the connected component of $S^2$ carved out by $\varphi\circ\mathcal{C}_j$ not containing $\varphi(\psi(\Omega))$. 

Now let $p\in B_r(0)\cap S^2 = W' \subset W$. Note, by definition, the curve $\partial\mathcal{S}^z\circ \mathcal{C}_j$, where as usual $\partial \mathcal{S}^z$ denotes the continuous extension of $\mathcal{S}^z\circ\psi^{-1}$ restricted to $\partial U$, does not intersect $W$ and hence $W'$. Therefore, we may, for each $j$, define $F_j^z\colon H_j(S^1\times[0,1])\longrightarrow S^2$
\begin{equation}
    F_j^z(H_j(s,t)) = \frac{(1-t)(\partial \mathcal{S}^z\circ\mathcal{C}_j)(s)-tp}{\lVert (1-t)(\partial\mathcal{S}^z\circ\mathcal{C}_j)(s)-tp \rVert}
\end{equation}
so that $F_j^z(H_j(S^1\times[0,1]))\cap W' = \varnothing$. By defining $F^z$ by $F^z\vert_{H_j(S^1\times[0,1])} = F_j^z$, we have constructed $(\varphi\circ\psi,F^z)$-extensions of $\mathcal{S}^z$ which satisfy the assumptions of theorem 3. $\blacksquare$

Corollary 4 ties our abstract mathematical results to the qualitative discussions presented in the main text, namely that as long as there exists some overlap of a given connected component as it transforms, then topological information can be relayed from one state to another. 

\section{Computing the Generalized Skyrmion number}

In this section, we discuss our approach to computing the generalized Skyrmion number from polarimetric measurements. To begin, we state our key steps.

\begin{enumerate}
    \item Determine the ``boundary'' of the field.
    \item Perform Gaussian process regression to obtain a smooth estimate of the field.
    \item Identify the connected components carved out by the boundary of the field. 
    \item For each connected component, pick a suitable form supported in it and determine an appropriate region of integration. Then perform numerical integration. 
\end{enumerate}
We now elaborate on the specifics of each step. \\

\noindent {\bf (1) Determining the ``boundary'' of the field}\\

As our field is generated by propagation through a gradient index lens, which is a confined medium, there is a natural boundary to consider at distance $z=0$, namely the boundary of the lens. However, as the field diffracts, this boundary is no longer unambiguous. The other natural boundary to consider in the context of free space propagation, at least in theory, is the boundary in the far field. In this case, it is the physical limitations of polarimetry, including the finite aperture of the camera and noise in measurements, that make capturing the polarization state on this boundary impractical. As a compromise between these two choices, we use the intensity profile of the beam, as shown in Fig.\ \ref{fig: Experiment} to determine an effective boundary in the following way. Notice from the figure that while the intensity profile of the beam spreads out in diffraction, there is still a clear transition as we move outwards from the center of the beam from light to dark, even at 2m. This observation gives rise to the following edge-detection-based algorithm for determining an effective boundary.

\begin{enumerate}
    \item Consider the intensity profile of the beam along various cross-sections in both the $x$ and $y$ directions.
    \item For each cross-section, perform edge-detection in the following way: (1) smooth the data using a Gaussian filter, (2) compute discrete derivatives using a Sobel filter, (3) thresholding to determine edges.
    \item Select the leftmost and rightmost edges for cross-sections in the $x$-direction, and the topmost and bottommost edges for cross-sections in the $y$-direction, these points then correspond to approximate points on the boundary. 
    \item Using the Coope method \citesupp{Coope}, fit a circle to the approximate boundary points. 
\end{enumerate}
The circle obtained by the algorithm above is then what we consider the ``boundary'' of the field.\\

\noindent {\bf (2) Gaussian process regression}\\

The next step is to obtain a smooth estimate of the field. This is important for two main reasons. Firstly, having an analytic expression for the field allows for more accurate computations of partial derivatives necessary in evaluating the generalized Skrymion number. Secondly, the analytic expression also allows the image of the boundary curve to be clear and well-defined, which is essential in determining the connected components that define the generalized Skyrmion number. In our work, we use radial basis functions as our kernel of choice.\\

\noindent {\bf (3) Identifying connected components}\\

The Gaussian process regression provides an analytic function that can be used to evaluate the boundary curve, which easily allows one to determine the connected components carved out by the boundary. There are, however, several technical points to note. As no polarimetry technique is noise-free and no beam generation technique is perfect, real-world limitations in experimentation can complicate the image of the boundary curve. Take, for example, a beam of uniform polarized light. Since, in practice, there will necessarily be fluctuations in the purity of the generated polarization state as well as noise in detection, the image of the predicted boundary curve will also fluctuate about some mean polarization state on the Poincar\'{e} sphere. This, in turn, potentially partitions the Poincar\'{e} sphere into one large component and several small components arising from self-intersections of the curve due to these fluctuations in the measured polarization state. While it is theoretically possible to compute the generalized Skyrmion number of these small components from the Gaussian process regression, the physical significance of these computed numbers is somewhat unclear, and we certainly do not expect them to be preserved in propagation. 

In the context of our experiments, the outer boundary of the curve loops around the Poincar\'{e} sphere twice due to the symmetry of the gradient index lens. However, because of the effect mentioned above, the curve occasionally self-intersects, producing a small component that technically has a generalized Skyrmion number. In order to overcome this, we adopt an area-based method to determine which components contribute to the generalized Skyrmion number, and which components do not. The method proceeds as follows.

\begin{enumerate}
    \item For each $p \in S^2$ not on the boundary curve, we give $p$ a score in the follow way. Let $\psi_p$ be the stereographic projection from $-p$ and $\partial \mathcal{S} \colon S^1 \longrightarrow S^2$ the boundary curve. Let 
    \begin{equation}
        r(p) = \sup 
    \{r\colon B_r(0)\cap \psi_p(\partial \mathcal{S}(S^1)) = \varnothing \}
    \end{equation}
    be the radius of the largest ball, after projection, that does not intersect the image of the boundary curve. We then set the score of $p$ to be the area of $\psi_p^{-1}(B_r(0))$ in $S^2$. We can estimate $r(p)$ numerically by taking the minimum distance of $\psi_p \circ \partial S$ from the origin evaluated at different values of $\theta \in [0, 2\pi]$. 
    \item By looping through points in $S^2$, we can determine the score of each point as defined above. We then consider connected components only if they contain a point with a score greater than a certain threshold. 
\end{enumerate}
In our case, the threshold we used was $\pi/4$, which is $1/16$ the total area of the sphere. While this threshold is somewhat arbitrary, it is the case, at least in our experiments, that the scores of different components are either much smaller than or much larger than this threshold. There is, however, significant scope for developing better algorithms for determining the generalized Skyrmion number. \\

\noindent {\bf (4) Numerical integration}\\

From the connected components, we compute the generalized Skyrmion number through numerical integration in the following way.
\begin{enumerate}
    \item Pick $p$ in the connected component of choice and compute $r(p)$.
    \item Define $\chi\colon S^2 \longrightarrow \mathbb{R}$
    \begin{equation*}
        \chi(s) = \left\{\begin{array}{r l}
            1, & \text{if $\psi_p(s)\in \overline{B_{0.95r(p)}}$,} \\
            0, & \text{otherwise}.
        \end{array}\right.
    \end{equation*}
    \item Compute $c = \int_{0}^{2\pi}\int_{0}^{\pi}\chi(\theta,\phi)\sin\theta d\theta d\phi$. Notice that this is just the area of a spherical cap and can be computed analytically. 
    \item Let $\mathcal{S}$ denote the approximation of the field obtained through Gaussian process regression. Plot $\chi(\mathcal{S})$ and determine sets $[r_1^{\text{min}}, r_1^{\text{max}}]\times[\theta_1^{\text{min}},\theta_1^{\text{max}}], \ldots, [r_k^{\text{min}}, r_k^{\text{max}}]\times[\theta_k^{\text{min}},\theta_k^{\text{max}}]$ so that 
    \begin{equation}
        \operatorname{supp} \chi \circ \mathcal{S} \subset \bigcup_{i=1}^{k}[r_i^{\text{min}}, r_i^{\text{max}}]\times[\theta_i^{\text{min}},\theta_i^{\text{max}}].
    \end{equation}
    \item For each $i=1, \ldots, k$, compute 
    \begin{equation}
        n_i = \int_{r_i^{\text{min}}}^{r_i^{\text{max}}}\int_{\theta_i^{\text{min}}}^{\theta_i^{\text{max}}}\chi(\mathcal{S})\mathcal{S}\cdot \left(\frac{\partial \mathcal{S}}{\partial r}\times \frac{\partial \mathcal{S}}{\partial \theta}\right) dr d\theta
    \end{equation}
    using a central difference for numerical derivatives and a $100\times 100$-point Gaussian quadrature rule for integration.
    \item Return $\frac{1}{c}\sum_{i=1}^k n_i$.
\end{enumerate}

The approach above is equivalent to taking the volume form $\omega = \chi\omega_0$, where $\omega_0$ is the standard normalized volume form on $S^2$ arising from its usual Riemannian metric. We pick $\omega$ in this way to ensure $c$ can be computed analytically and hence quickly and accurately, noting that while $\omega$ is not smooth, general density results guarantee the legitimacy of the computations. Step (4) helps reduce errors in numerical integration by excluding regions where the generalized Skyrmion density is 0, therefore narrowing down the sample points in step (5) to areas of interest. Since an analytic expression is used in integration, the procedure introduced is guaranteed to return an integer, with numerical errors resulting in deviations of the order $10^{-4}$. This error can be further reduced by simply increasing the number of sample points used in integration. Lastly, the usual Skyrmion number integral is also evaluated using $\mathcal{S}$ and a $100\times 100$-point Gaussian quadrature rule. 

\bibliographystylesupp{unsrt}
{\footnotesize \bibliographysupp{main}}

\renewcommand{\figurename}
{Extended Fig.}
\setcounter{figure}{0}
\section*{Extended Figures}
\vspace{-\baselineskip}
\begin{figure}[!ht]
\centering
\includegraphics[width=0.5\textwidth]{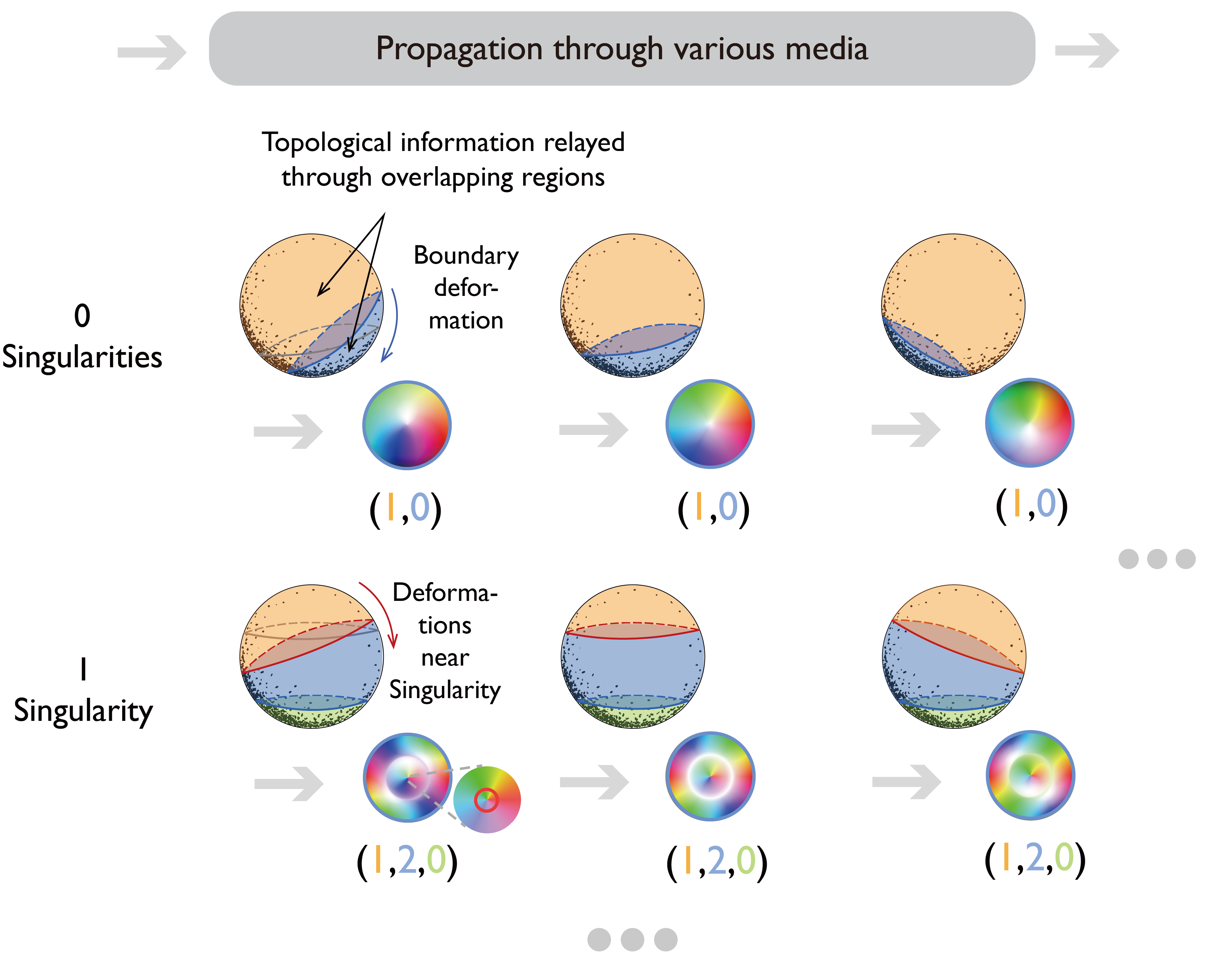}
\caption{\footnotesize {\bf Topological robustness of the generalized Skyrmion number.} Various fields in propagation and the corresponding components carved out by their boundaries. In each case, the smoothness of propagation gives rise to overlaps persistent throughout, which allows topological information to be relayed from the initial field to the final field. This then guarantees the preservation of the generalized Skyrmion number. The top row shows a scenario without singularities, and the bottom row a scenario with a single singularity. 
\label{fig: Concept4}}
\end{figure}

\newpage

\begin{figure}[!ht]
\centering
\includegraphics[width=0.5\textwidth]{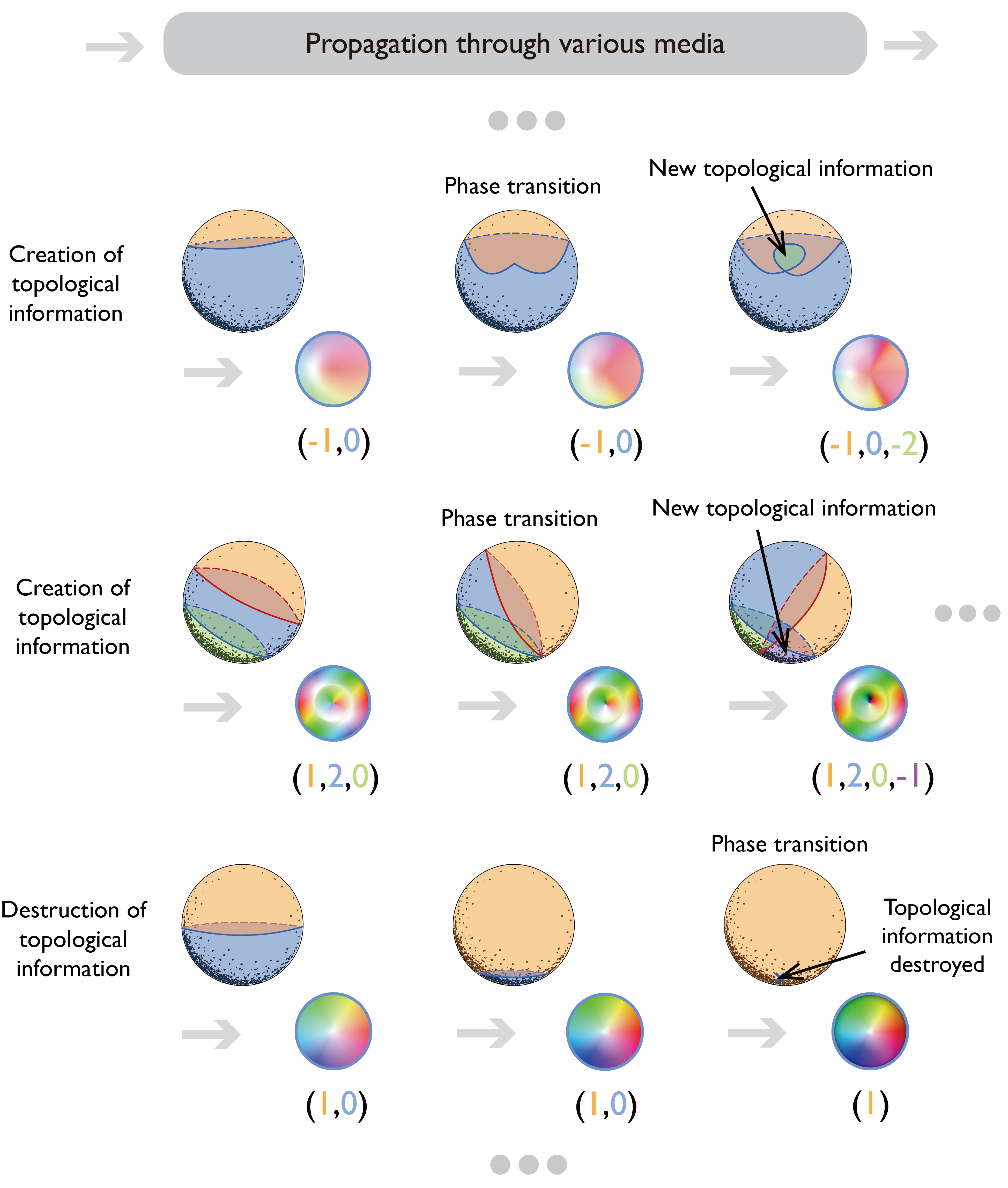}
\caption{\footnotesize {\bf Creation and destruction of topological information through phase transitions.} Various topological phase transitions that can occur which lead to the creation and destruction of topological information. (Top) Topological information is created through self-intersection of the outer boundary. (Middle) Topological information is created through the intersection of the outer boundary and the ``boundary'' of the singularity. (Bottom) Topological information is destroyed as the outer boundary collapses into a single point. 
\label{fig: Concept3}}
\end{figure}
\clearpage

\end{document}